\documentclass[conference]{IEEEtran}
\IEEEoverridecommandlockouts
\usepackage{cite}
\usepackage{amsmath,amssymb,amsfonts}
\usepackage{algorithmic}
\usepackage{graphicx}
\usepackage{subcaption}
\usepackage{textcomp}
\usepackage{xcolor}
\usepackage[a4paper, total={184mm,239mm}]{geometry}
\def\BibTeX{{\rm B\kern-.05em{\sc i\kern-.025em b}\kern-.08em
    T\kern-.1667em\lower.7ex\hbox{E}\kern-.125emX}}

\usepackage{lipsum}
\usepackage{enumitem}
\usepackage{bm}
\usepackage{multirow}
\usepackage{booktabs}
\usepackage{pifont}
\definecolor{cadmiumgreen}{rgb}{0.0, 0.42, 0.24}

\usepackage[ruled, vlined, norelsize]{algorithm2e}
\SetKwInput{KwInput}{Input}
\SetKwInput{KwOutput}{Output}
\SetKwInput{KwInit}{Initialization}
\SetKwInput{Kwprocedure}{Procedure}
\SetKwInput{KwParam}{Parameters}

\usepackage{tcolorbox}

\usepackage{tikz}
\usetikzlibrary{fit,calc}
\newcommand*{\tikzmk}[1]{\tikz[remember picture,overlay,] \node (#1) {};\ignorespaces}
\newcommand{\boxitonept}[1]{\tikz[remember picture,overlay]{\node[yshift=1pt,fill=#1,opacity=.25,fit={(A)($(B)+(.85\linewidth,.8\baselineskip)$)}] {};}\ignorespaces}
\newcommand{\boxitfourpt}[1]{\tikz[remember picture,overlay]{\node[yshift=4pt,fill=#1,opacity=.25,fit={(A)($(B)+(.85\linewidth,.8\baselineskip)$)}] {};}\ignorespaces}
\colorlet{pink}{red!40}

\def\todoen{1}
\if 1\todoen 
\newcommand\todo[1]{\textcolor{red}{\textbf{TODO:} #1}}
\else
\newcommand\todo[1]{}
\fi

\newcommand{\toolname}{\texttt{MABFuzz}}

\newcommand{\thehuzz}{\textit{TheHuzz}}

\newcommand{\rc}{{\tt Rocket Core}}
\newcommand{\boom}{{\tt BOOM}}
\newcommand{\cva}{{\tt CVA6}}

\usepackage{listings}

\definecolor{codegreen}{rgb}{0,0.6,0}
\definecolor{codegray}{rgb}{0.5,0.5,0.5}
\definecolor{codepurple}{rgb}{0.58,0,0.82}
\definecolor{backcolour}{rgb}{0.95,0.95,0.92}

\let\othelstnumber=\thelstnumber
\def\createlinenumber#1#2{
    \edef\thelstnumber{%
        \unexpanded{%
            \ifnum#1=\value{lstnumber}\relax
              #2%
            \else}%
        \expandafter\unexpanded\expandafter{\thelstnumber\othelstnumber\fi}%
    }
    \ifx\othelstnumber=\relax\else
      \let\othelstnumber\relax
    \fi
}

\lstdefinestyle{customc}{
  belowcaptionskip=1\baselineskip,
  breaklines=true,
  frame=single,
  xleftmargin=0.35cm,
  xrightmargin=0.15cm,
  numbers=left,
  numbersep=5pt,  
  language=C,
  showstringspaces=false,
  basicstyle=\footnotesize\ttfamily,
  keywordstyle=\bfseries\color{green!40!black},
  commentstyle=\itshape\color{purple!40!black},
  identifierstyle=\color{blue},
  stringstyle=\color{orange},
}

\lstdefinestyle{customcArianeExploit1}{
  breaklines=true,
  frame=single,
  xleftmargin=0.4cm,
  xrightmargin=0.2cm,
  numbers=left,
  numbersep=5pt,  
  language=C,
  showstringspaces=false,
  basicstyle=\footnotesize\ttfamily,
  keywordstyle=\bfseries\color{green!40!black},
  commentstyle=\itshape\color{purple!60!black},
  identifierstyle=\color{blue},
  stringstyle=\color{yellow!50!black},
  morekeywords={asm},
  keywordstyle=[2]\bfseries\color{brown!60!black},
}

\lstdefinestyle{customcArianeExploit}{
  breaklines=true,
  frame=single,
  xleftmargin=0.4cm,
  xrightmargin=0.2cm,
  numbers=left,
  numbersep=5pt,  
  language=C,
  showstringspaces=false,
  basicstyle=\footnotesize\ttfamily,
  keywordstyle=\bfseries\color{blue},
  commentstyle=\itshape\color{green!50!black},
  identifierstyle=\color{black},
  stringstyle=\color{brown},
  morekeywords={asm},
  keywordstyle=[2]\bfseries\color{black},
}

\lstdefinestyle{customlog}{
  breaklines=true,
  frame=single,
  xleftmargin=0.35cm,
  xrightmargin=0.15cm,
  numbers=left,
  numbersep=5pt,  
  language=C,
  showstringspaces=false,
  basicstyle=\footnotesize\ttfamily,
  keywordstyle=\color{blue},
  commentstyle=\itshape\color{purple!40!black},
  identifierstyle=\color{blue},
  stringstyle=\color{orange},
  keywords=[2]{INFO},
  keywords=[3]{ERROR},x
  keywordstyle=[2]\bfseries\color{green!40!black},
  keywordstyle=[3]\bfseries\color{red!500!black},
}

\definecolor{verilogcommentcolor}{RGB}{0,124,0}
\definecolor{verilogkeywordcolor}{RGB}{49,49,255}
\definecolor{backcolor}{RGB}{237,237,237}
\definecolor{verilogsystemcolor}{RGB}{128,0,255}
\definecolor{verilognumbercolor}{RGB}{255,143,102}
\definecolor{verilogstringcolor}{RGB}{160,160,160}
\definecolor{verilogdefinecolor}{RGB}{128,64,0}
\definecolor{verilogoperatorcolor}{RGB}{0,0,128}
\definecolor{pointcolor}{RGB}{192,0,0} 
\lstdefinestyle{prettyverilog}{
   backgroundcolor=\color{backcolor},
   language           = Verilog,
   commentstyle       = \color{verilogcommentcolor},
   alsoletter         = \$'0123456789\`,
   literate           = *{+}{{\verilogColorOperator{+}}}{1}%
                         {-}{{\verilogColorOperator{-}}}{1}%
                         {@}{{\verilogColorOperator{@}}}{1}%
                         {;}{{\verilogColorOperator{;}}}{1}%
                         {*}{{\verilogColorOperator{*}}}{1}%
                         {?}{{\verilogColorOperator{? }}}{1}%
                         {:}{{\verilogColorOperator{:}}}{1}%
                         {<}{{\verilogColorOperator{<}}}{1}%
                         {>}{{\verilogColorOperator{>}}}{1}%
                         {!}{{\verilogColorOperator{!}}}{1}%
                         {^}{{\verilogColorOperator{^}}}{1}%
                         {|}{{\verilogColorOperator{| }}}{1}%
                         {||}{{\verilogColorOperator{|| }}}{1}%
                         {=}{{\verilogColorOperator{= }}}{1}%
                         {==}{{\verilogColorOperator{== }}}{1}%
                         {=>}{{\verilogColorOperator{=> }}}{1}%
                         {[}{{\verilogColorOperator{[}}}{1}%
                         {]}{{\verilogColorOperator{]}}}{1}%
                         {(}{{\verilogColorOperator{(}}}{1}%
                         {)}{{\verilogColorOperator{)}}}{1}%
                         {rightbracket}{{\verilogColorOperator{)}}}{1}%
                         {,}{{\verilogColorOperator{,}}}{1}%
                         {.}{{\verilogColorOperator{.}}}{1}%
                         {~}{{\verilogColorOperator{$\sim$}}}{1}%
                         {\%}{{\verilogColorOperator{\%}}}{1}%
                         {\&}{{\verilogColorOperator{\& }}}{1}%
                         {\&\&}{{\verilogColorOperator{\&\& }}}{1}%
                         {\#}{{\verilogColorOperator{\#}}}{1}%
                         {\ /\ }{{\verilogColorOperator{\ /\ }}}{3}%
                         {\ _}{\ \_}{2}%
                        ,
   morestring         = [s][\color{verilogstringcolor}]{"}{"},%
   identifierstyle    = \color{black},
   vlogdefinestyle    = \color{verilogdefinecolor},
   vlogconstantstyle  = \color{verilognumbercolor},
   vlogsystemstyle    = \color{verilogsystemcolor},
   basicstyle         = \small\fontencoding{T1}\ttfamily,
  columns=fullflexible, 
   keywordstyle       = \bfseries\color{verilogkeywordcolor},
   morekeywords      = {val, when, port, coverage, unique},
   numbers            = left,
   numbersep          = 5pt,
   tabsize            = 2,
   escapeinside       = {/*!}{!*/},
   upquote            = true,
   sensitive          = true,
   showstringspaces   = false, 
   frame              = single, 
   breaklines         = true,
   abovecaptionskip   = 0pt,
   belowcaptionskip   = 0pt, 
   xleftmargin        =0.35cm,
   xrightmargin       =0.15cm,
   captionpos         = b,
   emph               = {Point, Point0, Point1, Point2, Point3, Point4, Point5, Point6, Point7, Point8, Point9},
   emphstyle          =\color{pointcolor},
   emph               = {[2] STVEC,SCOUNTEREN,MSTATUS,MTVEC,ML1_ICACHE_MISS,ML1_DCACHE_MISS,MITLB_MISS,MDTLB_MISS,
                             MLOAD,MSTORE,MEXCEPTION,MEXCEPTION_RET,MBRANCH_JUMP,MCALL,MRET,MMIS_PREDICT,MSB_FULL,
                             MIF_EMPTY,MHPM_COUNTER_17,MHPM_COUNTER_18,MHPM_COUNTER_19,MHPM_COUNTER_20,MHPM_COUNTER_21,
                             MHPM_COUNTER_22,MHPM_COUNTER_23,MHPM_COUNTER_24,MHPM_COUNTER_25,MHPM_COUNTER_26,MHPM_COUNTER_27,
                             MHPM_COUNTER_28,MHPM_COUNTER_29,MHPM_COUNTER_30,MHPM_COUNTER_31}, 
   emphstyle          = {[2]\bfseries\color{verilogkeywordcolor}}
}

\makeatletter

\newcommand\language@verilog{Verilog}
\expandafter\lst@NormedDef\expandafter\languageNormedDefd@verilog%
  \expandafter{\language@verilog}
  
\lst@SaveOutputDef{`'}\quotesngl@verilog
\lst@SaveOutputDef{``}\backtick@verilog
\lst@SaveOutputDef{`\$}\dollar@verilog

\newcommand\getfirstchar@verilog{}
\newcommand\getfirstchar@@verilog{}
\newcommand\firstchar@verilog{}
\def\getfirstchar@verilog#1{\getfirstchar@@verilog#1\relax}
\def\getfirstchar@@verilog#1#2\relax{\def\firstchar@verilog{#1}}

\newcommand\addedToOutput@verilog{}
\lst@AddToHook{Output}{\addedToOutput@verilog}

\newcommand\constantstyle@verilog{}
\lst@Key{vlogconstantstyle}\relax%
   {\def\constantstyle@verilog{#1}}

\newcommand\definestyle@verilog{}
\lst@Key{vlogdefinestyle}\relax%
   {\def\definestyle@verilog{#1}}

\newcommand\systemstyle@verilog{}
\lst@Key{vlogsystemstyle}\relax%
   {\def\systemstyle@verilog{#1}}

\newcount\currentchar@verilog
  
\newcommand\@ddedToOutput@verilog
{%
   \ifnum\lst@mode=\lst@Pmode%
      \expandafter\getfirstchar@verilog\expandafter{\the\lst@token}%
      \expandafter\ifx\firstchar@verilog\backtick@verilog
         \let\lst@thestyle\definestyle@verilog%
      \else
         \expandafter\ifx\firstchar@verilog\dollar@verilog
            \let\lst@thestyle\systemstyle@verilog%
         \else
            \expandafter\ifx\firstchar@verilog\quotesngl@verilog
               \let\lst@thestyle\constantstyle@verilog%
            \else
               \currentchar@verilog=48
               \loop
                  \expandafter\ifnum%
                  \expandafter`\firstchar@verilog=\currentchar@verilog%
                     \let\lst@thestyle\constantstyle@verilog%
                     \let\iterate\relax%
                  \fi
                  \advance\currentchar@verilog by \@ne%
                  \unless\ifnum\currentchar@verilog>57%
               \repeat%
            \fi
         \fi
      \fi
   \fi
}

\lst@AddToHook{PreInit}{%
  \ifx\lst@language\languageNormedDefd@verilog%
    \let\addedToOutput@verilog\@ddedToOutput@verilog%
  \fi
}

\newcommand{\verilogColorOperator}[1]
{%
  \ifnum\lst@mode=\lst@Pmode\relax%
   {\bfseries\textcolor{verilogoperatorcolor}{#1}}%
  \else
    #1%
  \fi
}

\makeatother

\lstdefinestyle{mystyle}{
    commentstyle=\textit,
    keywordstyle=\textbf,
    stringstyle=\color{codepurple},
    basicstyle=\ttfamily,
    breakatwhitespace=false,         
    breaklines=true,      
    frame=single, 
    framexleftmargin=\parindent,
    captionpos=b,                    
    keepspaces=true,                 
    numbers=left,    
    numberstyle=\normalsize,
    stepnumber=1,
    numbersep=5pt,   
    xleftmargin=1.5\parindent,
    showspaces=false,                
    showstringspaces=false,
    showtabs=false,                  
    tabsize=2
}

\usepackage[normalem]{ulem}  
\def\removesen{1}
\def\adden{1}

\if 1\removesen
\newcommand\red[1]{\textcolor{red!40!white}{\sout{#1}}}
\else
\newcommand\red[1]{}
\fi

\if 1\adden

\else

\fi

\begin{document}

\title{\toolname: Multi-Armed Bandit Algorithms for Fuzzing Processors\\
}

\author{
\IEEEauthorblockN{Vasudev Gohil$^{\ast,\dagger}$\thanks{$^\ast$These authors contributed equally to this work.}, Rahul Kande$^{\ast,\dagger}$, Chen Chen$^\dagger$, Ahmad-Reza Sadeghi$^\ddagger$, and Jeyavijayan Rajendran$^\dagger$}
\IEEEauthorblockA{$^\dagger$Texas A\&M University, $^\ddagger$Technische Universit\"at Darmstadt}
\tt{$^\dagger$\{gohil.vasudev, rahulkande, chenc, jv.rajendran\}@tamu.edu,}\\
\tt{$^\ddagger$\{ahmad.sadeghi\}@trust.tu-darmstadt.de}
}

\maketitle

\begin{abstract}
As the complexities of processors keep increasing, the task of effectively verifying their integrity and security becomes ever more daunting. The intricate web of instructions, microarchitectural features, and interdependencies woven into modern processors pose a formidable challenge for even the most diligent verification and security engineers. To tackle this growing concern, recently, researchers have developed fuzzing techniques explicitly tailored for hardware processors. However, a prevailing issue with these hardware fuzzers is their heavy reliance on static strategies to make decisions in their algorithms. 
To address this problem, we develop a novel dynamic and adaptive decision-making framework,~\toolname{}, that uses multi-armed bandit (MAB) algorithms to fuzz processors. \toolname{} is agnostic to, and hence, applicable to, any existing hardware fuzzer. In the process of designing~\toolname, we encounter challenges related to the compatibility of MAB algorithms with fuzzers and maximizing their efficacy for fuzzing. We overcome these challenges by modifying the fuzzing process and tailoring MAB algorithms to accommodate special requirements for hardware fuzzing.

We integrate three widely used MAB algorithms in a state-of-the-art hardware fuzzer and evaluate them on three popular RISC-V-based processors. Experimental results demonstrate the ability of~\toolname{} to cover a broader spectrum of processors' intricate landscapes and doing so with remarkable efficiency. 
In particular,~\toolname{} achieves up to $\mathbf{308}\times$ speedup in detecting vulnerabilities and up to $\mathbf{5}\times$ speedup in achieving coverage compared to a state-of-the-art technique.
\end{abstract}


\section{Introduction}\label{sec:introduction}

Modern processors are getting increasingly complex with the goal of offloading software functionalities such as cryptographic evaluations to hardware, all while optimizing power, performance, and area. 
This complexity, paired with the shrinking verification time, poses an immense challenge to existing verification tools.
Insufficient verification will lead to a rise in security-critical hardware vulnerabilities that can be exploited by cross-layer attacks, threatening the reputation of companies, and undermining the safety, security, and resilience of critical infrastructure~\cite{dessouky2019hardfails}. 
For example, the broken hyper-threading vulnerability in Intel Skylake and Kaby Lake leads to data corruption, or data loss~\cite{intelhyper}.
The \textit{Zenbleed} vulnerability in AMD Zen 2, allows attackers to potentially access secret data~\cite{zenbleed}.
Billions of dollars are being invested by both industry and government entities in research and development efforts aimed at fortifying the security posture of these critical components of modern computing infrastructure.

\subsection{Hardware Fuzzing}
One such approach that has gained prominence is hardware fuzzing. 
Hardware fuzzers employ automated testing techniques to probe processors for vulnerabilities. 
Several hardware fuzzers have already been developed~\cite{muduli2020hyperfuzzing,hur2021difuzzrtl,fuzzhwlikesw,kande2022thehuzz,chen2023hypfuzz,ragab_bugsbunny_2022,chen2023psofuzz}. These hardware fuzzers have proven to be faster, automated, and scalable compared to traditional hardware verification techniques such as random regression, directed testing, or formal verification.
Even commercial enterprises such as Intel and Google are actively developing new and efficient hardware fuzzers for vulnerability detection~\cite{presifuzz,fuzzhwlikesw}.

\subsection{Limitations of Existing Hardware Fuzzers}
Although hardware fuzzing is promising, existing hardware fuzzers are inefficient in verifying processors because they make many static decisions disregarding the design complexity and the design space explored. 
Recent research has shown that replacing the static decisions in the mutation engine of fuzzer with a dynamic strategy is likely to cover more points in the design~\cite{chen2023psofuzz}. 
However, fuzzers make many other static decisions that are not under the scope of~\cite{chen2023psofuzz}. 
For example, \cite{kande2022thehuzz} selects the tests from its database in a static first-in-first-out method and does not prioritize selecting the tests with more potential first.
To address this limitation of static strategies in fuzzers, we develop an approach to equip any hardware fuzzer with a dynamic decision-making technique, multi-armed bandit (MAB) algorithms. 

MAB algorithms are devised for dynamic decision-making under uncertainty and striking a good balance between the exploration of novel decisions and exploitation of known, well-performing decisions~\cite{sutton2018reinforcement}. Researchers have used MAB algorithms to develop promising solutions to several security problems such as intrusion detection~\cite{tariq2022network} and securing cyber-physical systems~\cite{ferdowsi2019cyber}. However, to the best of our knowledge, the potential of MAB algorithms in hardware fuzzing has not been explored. Through this work, we aim to fill this gap by using MAB algorithms to optimize the selection of input tests and speed up vulnerability detection.

\subsection{Our Contributions}
Applying MAB algorithms to hardware fuzzing is not trivial. 
This is because new coverage achieved by fuzzers decreases with time and MAB algorithms are not designed to work in this setting where the returns diminish with time. So, to ensure that our technique is efficient and effective, we need to (i)~devise a way to monitor the coverage returns from the fuzzer, and (ii)~modify MAB algorithms so that they are compatible with this diminishing returns property of hardware fuzzers.
We designed \toolname{} by addressing these challenges which resulted in faster vulnerability detection speed and covered more hardware. 
Overall, the main contributions of this work are as follows:
\begin{enumerate}
    \item To the best of our knowledge, we develop the first technique that uses MAB algorithms, \toolname{}, to select test inputs in hardware fuzzers.
    \item We overcome challenges in adapting MAB to hardware fuzzers. In particular, we develop monitors to identify saturated inputs and modify MAB algorithms to handle such test inputs.
    \item We integrate three widely-used MAB algorithms in a hardware fuzzer, demonstrating the agnostic nature of our technique.
    \item We evaluate \toolname{} on three widely-used, open-sourced RISC-V processors and achieve up to $308.89\times$ speedup in detecting vulnerabilities and up to $5.38\times$ speedup in achieving coverage compared to the state-of-the-art simulation-based fuzzer.
\end{enumerate}

\section{Background}\label{sec:background}

\subsection{Hardware Processor Fuzzers}\label{sec:fuzzing_background}

Hardware fuzzing is a regression-based vulnerability detection technique that tests the target hardware, such as a processor, by iteratively generating inputs, termed \textit{tests} (i.e., a sequence of instructions or binary executables) and executing on the target~\cite{kande2022thehuzz,chen2023hypfuzz,chen2023psofuzz}. 
It generates the first set of these inputs, termed \textit{seeds} by randomly creating instructions.
These seeds populate the fuzzer's input database, i.e., \textit{test pool}. 
Fuzzer selects \textit{tests} from the input pool, simulates the target to collect coverage feedback, and trace log outputs.
The coverage feedback comprises information about the activity in hardware such as toggling a register bit or entering a control path in the form of coverage \textit{points}. 
Fuzzer \textit{mutates} tests (a.k.a \textit{interesting tests}) that cause new activity in the hardware, i.e., cover new points. 
This mutation performs bit manipulation operations such as flipping a bit on the current test to create new tests and adds them to the input pool.  
Most processor fuzzers use differential testing to detect vulnerabilities~\cite{hur2021difuzzrtl,kande2022thehuzz,chen2023hypfuzz,chen2023psofuzz}. 
This technique compares the changes in the architectural state of the hardware processor and a reference model for every executed instruction to detect mismatches and flag potential vulnerabilities. 
They use an ISA simulator, such as \texttt{SPIKE}~\cite{spike} for RISC-V ISA~\cite{riscv_home}, as the reference model. This reference model is run with the same input as the hardware processor. Hence, it generates the expected correct architectural state to compare with the state of the hardware processor.

\subsection{Multi-Armed Bandits (MABs)}
MABs represent a framework for decision-making under uncertainty. In a typical MAB problem, an agent is faced with a set of choices to take, or \textit{arms} to pull, each of which yields an uncertain reward. The agent's challenge lies in balancing the exploration of these arms to learn their reward probabilities and the exploitation of known information to maximize cumulative rewards. 
This exploration-exploitation trade-off is pervasive in numerous real-world applications, including online advertising, clinical trials, and recommendation systems.

There has been extensive research on the MAB problem, resulting in a diverse range of algorithms.
These algorithms aim to optimize decision-making strategies based on the evolving understanding of arms' performances.
Researchers have used MABs and its more general form, reinforcement learning, to develop promising solutions for problems in hardware security~\cite{gohil2022deterrent,chen2023adatest,gohil2022attrition,guo2022vulnerability}. However, to the best of our knowledge, this is the first work to use MAB algorithms for finding vulnerabilities in processors using fuzzing.

\section{\toolname: Multi-Armed Bandit Algorithms for Fuzzing Processors}

In this section, we first describe why MAB algorithms are a good fit for finding vulnerabilities using hardware fuzzing. Then, we detail our preliminary formulation. Next, we explain some challenges the preliminary formulation faces and our solutions. Finally, we detail our final~\toolname{} formulation.

\subsection{Why MAB Algorithms?}
As described in Sec.~\ref{sec:fuzzing_background}, hardware fuzzing involves several moving pieces, e.g., 
generating seeds, selecting tests to simulate, prioritizing interesting tests to mutate to obtain high coverage, deciding mutation operators to apply to these tests, and so on. Most existing processor fuzzers either make these decisions randomly or use static strategies~\cite{kande2022thehuzz,hur2021difuzzrtl,chen2023hypfuzz}. However, recent research has shown this is not ideal. For instance, dynamically selecting mutation operators is likely to cover more design points than static probabilities~\cite{chen2023psofuzz}. A hardware fuzzer should have two key features to obtain higher coverage and find vulnerabilities faster, as explained next.

\noindent\textbf{Dynamic Decision-Making Under Uncertainty.} 
An effective hardware fuzzer must continually adapt its input generation strategies to explore untested design regions (which also change with time)  within the hardware's logic and functionalities. 
It must make dynamic decisions when faced with the ambiguity of which tests are most likely to unveil vulnerabilities.

\noindent\textbf{Balanced Trade-Off Between Exploration and Exploitation.} Another feature required from a good hardware fuzzer is its ability to strike the delicate equilibrium between exploration and exploitation, which is driven by the inherent constraints of limited resources. 
The fuzzer must judiciously allocate its constrained resources, such as time or number of tests, to explore new areas of the design while capitalizing on what is already known, effectively maximizing coverage. This careful balance ensures not only the discovery of hidden vulnerabilities but also the efficient utilization of scarce resources.

Tasks that require these features are the kind of problems that MAB algorithms are ideal for solving---\textbf{sequential decision-making under uncertainty} and striking a \textbf{balance between exploration and exploitation} in a search space.
Hence, next, we architect a preliminary framework to fuzz processors with MAB algorithms to find vulnerabilities.

\subsection{Preliminary Formulation}
There are several avenues for decision-making in hardware fuzzers where MAB algorithms can be applied. 
One of the most critical points of these avenues is deciding which seeds to pick
from a pool of seeds. 
Most existing hardware fuzzers select seeds uniformly at random
or select seeds greedily based on their historical coverage~\cite{kande2022thehuzz,chen2023hypfuzz}.
This is not ideal, as different seeds can have vastly different impacts on covering new points in hardware. The following example demonstrates this.

\begin{lstlisting}[language=Verilog, label = {listing:example_listing}, caption={Code snippet for motivational example},style=prettyverilog,float,belowskip=-5pt,aboveskip=20pt,firstnumber=1,linewidth=\linewidth]
reg MyReg1;
reg MyReg2;
. . .
always@(posedge clkrightbracket begin
    if (MyReg1rightbracket
        cov1 // coverage point 1
    if (!MyReg1rightbracket
        cov2  // coverage point 2
    if (MyReg2rightbracket
        cov3 // coverage point 3
    . . .
end
\end{lstlisting}
\noindent\textbf{Motivational Example.} Consider the code shown in Listing~\ref{listing:example_listing} with three target coverage points, \texttt{cov1}, \texttt{cov2}, and \texttt{cov3}.
Next, suppose we have two seeds, $S_1$ and $S_2$, such that $S_1$ contains instructions that control the values of \texttt{MyReg1}, and $S_2$ contains instructions that control the value of \texttt{MyReg2}.
Also, suppose that in the past, $S_1$ covered \texttt{cov1} and \texttt{cov2}, and $S_2$ has not been selected so far.
If we pick seeds greedily (i.e., exploit seeds that have performed well in the past), we would pick $S_1$ and not cover \texttt{cov3}.
On the other hand, if we use a dynamic strategy that balances well between the exploitation of known information (i.e., select seeds that have performed well historically) and exploration for updating information (i.e., try out untested or less tested seeds), we would likely select $S_2$, resulting in covering a novel region of the design, \texttt{cov3}.
Note that although this is a hypothetical example, the implications are applicable in the case of real-world processors, too, as evidenced by the V7 vulnerability in our results in Sec.~\ref{sec:vulnerability_detection_results}.

We now map the seed selection problem in hardware fuzzing to an MAB problem by defining the different aspects of MABs in terms of hardware fuzzing.

\begin{itemize}[leftmargin=*]
\item \textbf{Arms $\mathcal{A}$} is the set of arms from which the agent can choose. Each individual arm, $a_i \in \mathcal{A}$, corresponds to a different seed in the set of seeds of the fuzzer. 
Note that at each time step, $t$, the agent can only select, i.e., pull, one arm.

\item \textbf{Reward Distributions $\mathcal{R}$} indicate the likelihood of rewards for the arms in $\mathcal{A}$.
For hardware fuzzers, the reward for pulling an arm, i.e., selecting a seed, changes with time. So, we denote the reward for pulling an arm $a_i$ at time $t$, as $R_t(a_i)$. Since the objective is to maximize coverage, we define this reward as:
\begin{equation*}
    \begin{aligned}
        R_t(a_i) &= \alpha \times |cov^{L}_{t}(a_i)| + (1-\alpha) \times |cov^{G}_{t}(a_i)| \text{, where}\\
        cov^{L}_{t}(a_i) &= \{c_j \mid c_j \text{ is covered by } a_i \text{ at } t \text{ but not before } t\}\\
        cov^{G}_{t}(a_i) &= cov^{L}_{t}(a_i) \cap \{ c_j \mid c_j \text{ is not covered before } t\}
    \end{aligned}
\end{equation*}

Here, $cov^{L}_{t}(a_i)$ denotes the set of coverage points that are covered by $a_i$ in the current step, $t$, but have not been covered by $a_i$ in the past (superscript $L$ denotes local).
$cov^{G}_{t}(a_i)$ denotes the set of coverage points that are covered by $a_i$ in the current step, $t$, and have not been covered by any test of any arm in the past (superscript $G$ denotes global). Thus, the reward, $R_t(a_i)$, is a weighted sum of the number of global and local new coverage points. The parameter $\alpha \in [0,1]$ controls the weight, i.e., the relative importance of finding new coverage points not covered by arm $a_i$ and those not covered by any arm. 
\end{itemize}

\begin{figure}
    \centering
    \includegraphics[width=0.5\textwidth,trim={0.5cm 0.5cm 0.5cm 0.5cm},clip]{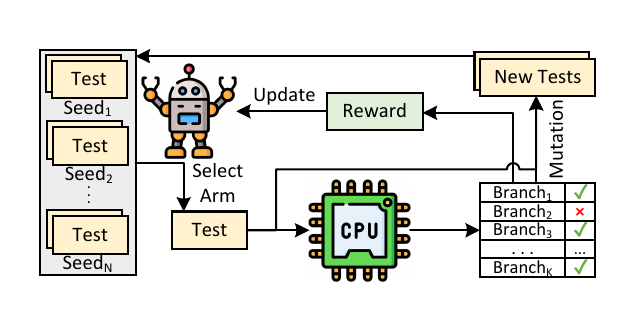}
    \caption{High-level flow of the preliminary formulation.}
    \label{fig:flow}
     \vspace{0.5cm}
\end{figure}
Fig.~\ref{fig:flow} illustrates the high-level flow of this preliminary formulation. 
First, a set of seeds, with one seed for each arm, is created. 
Each arm has a test pool consisting of its tests. 
Then, the agent picks one of the arms according to the initial arm selection probabilities
of an MAB algorithm.
Then, a test from the selected arm's test pool is simulated to obtain coverage information. Next, based on the coverage information, the original test is mutated to generate new tests which are added to the arm's pool. Additionally, the coverage information is also used to provide the reward to the agent for the chosen arm. The MAB algorithm uses the obtained reward to update the arm selection probabilities. Then, the agent again selects an arm according to the updated arm selection probabilities, and the cycle continues. Through a balance of exploration and exploitation, the agent tries to maximize the design coverage.

Since the application of MAB for hardware fuzzing is unexplored, we designed~\toolname{} to support any MAB algorithm. In our experiments, we use three different widely-used MAB algorithms, $\varepsilon$-greedy~\cite{sutton2018reinforcement}, UCB~\cite{auer2002finite_ucb1}, and EXP3~\cite{auer2002nonstochastic_exp3}, that have achieve good performance historically.

\begin{figure*}[t]
    \centering
    \includegraphics[width=\textwidth,trim={0.5cm 0.5cm 0.5cm 0.5cm},clip]{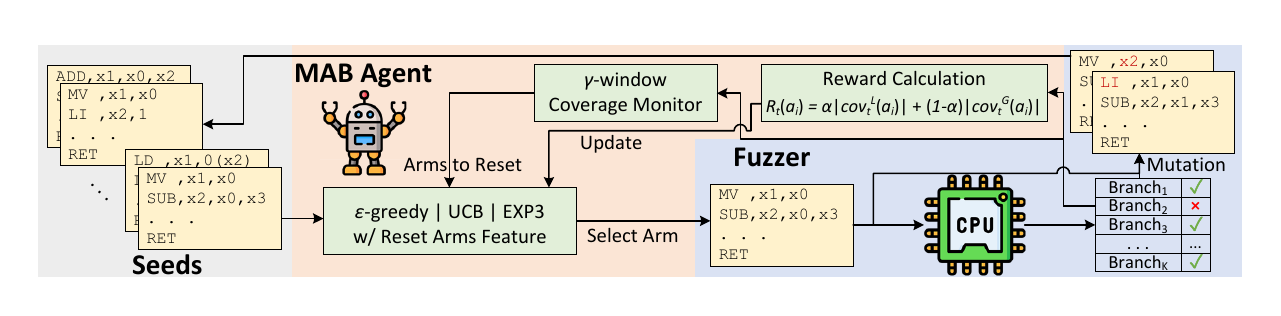}
    \caption{Final~\toolname{} framework.}
    \label{fig:final_framework}
\end{figure*}

\SetArgSty{textnormal}
\newcommand\mycommfont[1]{\footnotesize\ttfamily\textcolor{verilogcommentcolor}{#1}}
\SetCommentSty{mycommfont}
\setlength{\textfloatsep}{2pt}
\begin{algorithm}[t]
\LinesNumbered
\DontPrintSemicolon
\KwInit{$Q(a)\leftarrow 0$, $N(a)\leftarrow 0$  $\forall a \in \mathcal{A}; t\leftarrow0$}
\While{Fuzzing continues}{
    $t\leftarrow t + 1$\\
    \uIf{algorithm == $\varepsilon$-greedy}{
          $A\leftarrow 
          \begin{cases}
          {\arg\max_a } Q(a) &\text{ with probability } 1-\varepsilon \\
          \text{random } a\in\mathcal{A} &\text{ with probability } \varepsilon 
          \end{cases}
          $\\
    }
    \ElseIf{algorithm == UCB}{
        $A\leftarrow {\arg\max_a } \left[Q(a) + \sqrt{\frac{2\times\text{ln}t}{N(a)}}\right]$\\
    }
    $R_t(A) \leftarrow PullArm(A)$ \tcp*[l]{Simulate Tests}
    $N(A) \leftarrow N(A) + 1$\\
    $Q(A) \leftarrow Q(A) + \frac{1}{N(A)}\left[R_t(A)-Q(A)\right]$\\
    \tikzmk{A}
    \If{$A$ needs to be reset}{
        $N(A) \leftarrow 0$\\
        $Q(A) \leftarrow 0$
    }
    \tikzmk{B}
    \boxitonept{pink}
    \vspace{-.5\baselineskip}
}
\caption{Modified $\varepsilon$-greedy~\cite{sutton2018reinforcement} and UCB~\cite{auer2002finite_ucb1}}\label{alg:epsilon_greedy_and_ucb}
\end{algorithm}

\subsection{Tackling Diminishing Marginal Rewards}
The preliminary formulation described above works to an extent, but a crucial property of hardware fuzzing is not taken into account by the MAB algorithms: after a few time steps or iterations, the number of new coverage points obtained by mutating tests decreases drastically with time. 
To address this, we (i)~modify the fuzzing loop to check when an arm is no longer effective and (ii)~modify the MAB algorithms to accommodate this peculiar property of hardware fuzzing.

\noindent\textbf{Recognizing and Handling Saturated Arms.} 
To identify when an arm is not capable of covering novel regions of the design, we monitor the coverage of the arm over the most recent window of $\gamma$ iterations when the arm was picked by the agent. If the coverage does not increase in that $\gamma$-window, we mark the arm as depleted and replace it with a new arm.\footnote{ 
$\gamma$ controls the trade-off between possibly covering deeper regions of the design at the cost of time (large $\gamma$) versus trying to explore other novel regions in the design while potentially sacrificing deep coverage points (small $\gamma$).} 
We call this \textit{resetting} an arm. However, resetting an arm leads to another complication: the existing MAB algorithms are not designed to work when an arm is reset, i.e., replaced with another new arm. Next, we address this complication.

\noindent\textbf{Accommodating Reset Arms in MAB Algorithms.} The core idea to accommodate a reset arm is to treat the arm as a fresh arm and reset appropriate terms (e.g., the number of times the arm has been picked in the $\varepsilon$-greedy algorithm). 
Algorithms~\ref{alg:epsilon_greedy_and_ucb} and~\ref{alg:exp3} detail our modified MAB algorithms for $\varepsilon$-greedy and UCB, and EXP3 algorithms, respectively.
The parts highlighted in red are our modifications to accommodate reset arms.\footnote{Since the non-highlighted parts of the algorithms are unchanged, we refer an interested reader to~\cite{sutton2018reinforcement,auer2002finite_ucb1,auer2002nonstochastic_exp3} for their explanations.} For $\varepsilon$-greedy and UCB, if the selected arm, $A$, needs to be reset, we reset the counter for the number of times the arm has been pulled, $N(A)$, and its value, $Q(A)$ to $0$ (lines 11 and 12 in Algorithm~\ref{alg:epsilon_greedy_and_ucb}). For EXP3, 
we set the weight of the arm, $W(A)$, 
as the average weight of the other arms (line 10 in Algorithm~\ref{alg:exp3}). Additionally, since EXP3 requires normalization, we divide the reward by the total number of coverage points, $|C|$ (line 6).

\begin{algorithm}[t]
\LinesNumbered
\DontPrintSemicolon
\KwParam{$\eta \in (0,1]$ \tcp*[l]{learning rate}} 
\KwInit{$W(a)\leftarrow 1$ $\forall a \in \mathcal{A}; t\leftarrow0$}
\While{Fuzzing continues}{
    $t\leftarrow t + 1$\\
    $P(a)\leftarrow (1-\eta) \frac{W(a)}{\sum_{j\in\mathcal{A}}W(j)} + \frac{\eta}{|\mathcal{A}|} \quad \forall a\in \mathcal{A}$\\
    $A \sim P$ \tcp*[l]{Randomly sample $A$ according to $P$}
    $R_t(A) \leftarrow PullArm(A)$ \tcp*[l]{Simulate Tests}
    \tikzmk{A}
    $R_t(A) \leftarrow \frac{R_t(A)}{|C|}$ \tcp*[l]{Normalize reward}
    \tikzmk{B}
    \boxitfourpt{pink}
    $x\leftarrow \frac{R_t(A)}{P(A)}$\\
    $W(A) \leftarrow W(A)\times e^{(\eta x/|\mathcal{A}|)}$\\
    \tikzmk{A}
    \If{$A$ needs to be reset}{
        $W(A) \leftarrow \frac{\sum_{j\in \mathcal{A}\setminus A}{W(j)}}{|\mathcal{A}|-1}$ \tcp*[l]{Set weight of $A$ as average weight of other arms}
    }
    \tikzmk{B}
    \boxitonept{pink}
    \vspace{-.5\baselineskip}
}
\caption{Modified EXP3~\cite{auer2002nonstochastic_exp3}}\label{alg:exp3}
\end{algorithm}

\subsection{Final \toolname{} Formulation}
Fig.~\ref{fig:final_framework} illustrates the final~\toolname{} framework. Given a seed pool with each seed corresponding to an arm, the MAB agent selects an arm. This selection is done according to a predetermined modified MAB algorithm with the reset arms feature, such as the modified UCB algorithm (Algorithm~\ref{alg:epsilon_greedy_and_ucb}) or the modified EXP3 algorithm (Algorithm~\ref{alg:exp3}). Next, the test corresponding to the selected arm is simulated on the target processor design to obtain the coverage information. After that, the original test is mutated to generate new tests, which are added to the pool, and the coverage information is 
(i)~used to compute the reward for the agent's choice of arm and (ii)~monitored to decide if the selected arm needs to be reset. The computed reward is used to update the MAB agent, which again selects an arm, and the cycle continues.

\section{Experimental Evaluation}\label{sec:experimental_evaluation}

\begin{figure*}[htb!]
    \centering
    \includegraphics[width=\textwidth]{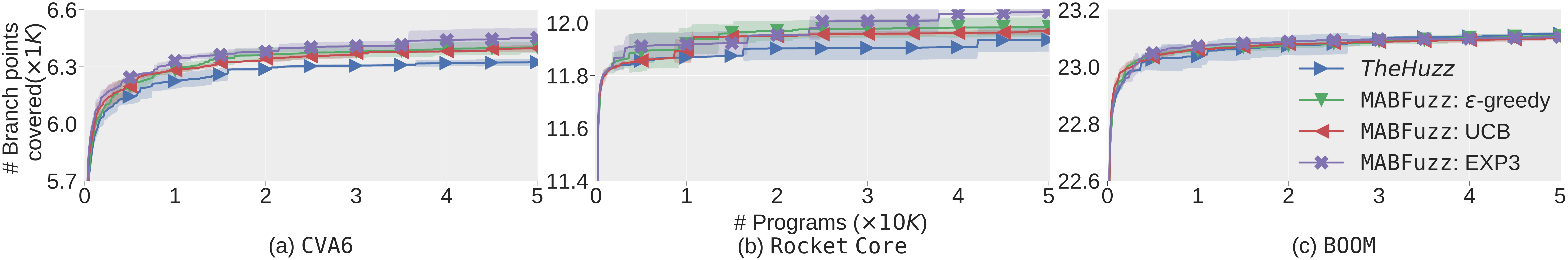}
    \caption{Branch coverage achieved by \toolname: $\varepsilon$-greedy, \toolname: UCB, and \toolname: EXP3 compared to~\thehuzz~\cite{kande2022thehuzz}.}
    \label{fig:covPlot}
\end{figure*}

\begin{figure*}[!t]
    \centering
    \includegraphics[width=\textwidth,trim={0cm 0.3cm 0cm 0.2cm},clip]{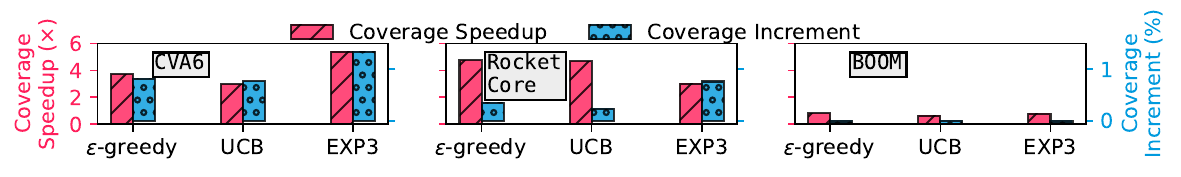}
    \caption{Coverage speedup and increment compared to the base fuzzer,~\thehuzz~\cite{kande2022thehuzz}, for~\toolname: $\varepsilon$-greedy, UCB, and EXP3.}
    \label{fig:covRate}
\end{figure*}

\subsection{Experimental Setup}
\noindent\textbf{Implementation Setup.} 
We implement~\toolname{} with~\thehuzz{} as the base fuzzer since~\thehuzz{} is one of the state-of-the-art simulation-based processor fuzzers~\cite{kande2022thehuzz}.
Similar to~\cite{kande2022thehuzz}, we implement~\thehuzz{} and~\toolname{} in Python. 
We run all experiments on a machine with CentOS Linux distribution on Intel Xeon processors with $64$ threads, $512$GB RAM, and a $2.6$GHz clock. 
We set the number of arms for~\toolname{} as $10$. As the ultimate objective is to maximize coverage, we set the parameter $\alpha$ (that controls the relative weights of new local and global coverage points) as $0.25$, meaning that we assign $3\times$ importance to a point previously uncovered by any arm compared to a point previously uncovered by the chosen arm but covered by some other arm. Based on experiments, we observed that the value of $3$ for the reset threshold, $\gamma$, yields good results, so we set it as $3$. We set the learning rate, $\eta$, in Algorithm~\ref{alg:exp3} as $0.1$.

\noindent\textbf{Evaluation Setup.} Following prior works, we evaluate~\toolname{} on three open-source processors, \cva~\cite{cva6}, \rc~\cite{rocket_chip_generator}, and \boom~\cite{boom}, since most commercial processors are close-sourced~\cite{kande2022thehuzz}. These processors have various features such as out-of-order execution and a custom single instruction-multiple data floating point unit in \cva, in-order execution in \rc, and \boom{} is a superscalar processor. Additionally, although these processors are open-sourced, they can fully boot the Linux operating system, demonstrating their practicality. We simulate the processors using Synopsys VCS and use Chipyard as the system-on-chip simulation environment. We ran experiments with $50,000$ tests for each benchmark for all fuzzers and repeated each experiment at least three times to reduce randomness in results. We use the branch coverage as the metric for comparison because it is highly correlated with vulnerability detection~\cite{mockus2009test}.

\begin{table}[!t]
\caption{Vulnerability detection speedup compared to \thehuzz{}~\cite{kande2022thehuzz}. Vulnerabilities V1---V6 are for \cva~\cite{cva6} and V7 is for \rc~\cite{rocket_chip_generator}.}
\label{tab:b_list_v1}
\resizebox{0.49\textwidth}{!}{%
\begin{tabular}{cccccc}
\toprule
\multirow{2}{*}{Vulnerability} & \multirow{2}{*}{CWE \#} & \thehuzz{}~\cite{kande2022thehuzz} & $\varepsilon$-greedy & UCB & EXP3 \\ \cmidrule(lr){3-3} \cmidrule(lr){4-4} \cmidrule(lr){5-5} \cmidrule(lr){6-6}
 &  & \# Tests & Speedup & Speedup & Speedup \\ \midrule 
\begin{tabular}[c]{@{}c@{}}V1: \texttt{FENCE.I} instruction\\decoded incorrectly\end{tabular} & 440 & $6.00\times10^{2}$ & $1.56\times$ & $\mathbf{13.04}\times$ & $4.5\times$ \\ \cmidrule(lr){1-6}
\begin{tabular}[c]{@{}c@{}}V2: Some \textit{illegal} instructions\\ can be executed\end{tabular} & 1242 & $1.48\times10^3$ & $0.73\times$ & $2.26\times$ & $\mathbf{8.33}\times$ \\ \cmidrule(lr){1-6}
\begin{tabular}[c]{@{}c@{}}V3: Exception type incorrectly\\ propagated in instruction queue\end{tabular} & 1202 & $2.39\times10^2$ & $\mathbf{59.75}\times$ & $7.71\times$ & $11.95\times$ \\ \cmidrule(lr){1-6}
\begin{tabular}[c]{@{}c@{}}V4: Undetected cache\\coherency violation\end{tabular} & 1202 & $1.20\times10^3$ & $\mathbf{2.43}\times$ & $1.48\times$ & $1.22\times$ \\ \cmidrule(lr){1-6}
\begin{tabular}[c]{@{}c@{}}V5: Exception not thrown\\when \textit{invalid} addresses accessed
\end{tabular} & 1252 & $2.50\times10^0$ & $0.35\times$ & $0.13\times$ & $\mathbf{0.63}\times$ \\ \cmidrule(lr){1-6}
 \begin{tabular}[c]{@{}c@{}}V6: Accessing unimplemented\\CSRs returns X-values\end{tabular} & 1281 & $1.41\times10^2$ & $2.33\times$ & $2.11\times$ & $\mathbf{2.36}\times$ \\ \cmidrule(lr){1-6}
\begin{tabular}[c]{@{}c@{}}V7: \texttt{EBREAK} does not\\ increase instruction count\end{tabular} & 1201 & $9.27\times10^2$ & $\mathbf{308.89}\times$ & $185.34\times$ & $73.16\times$ \\ 
\bottomrule
\end{tabular}%
}
\end{table}

\subsection{Vulnerability Detection}\label{sec:vulnerability_detection_results} 
\toolname{} uses the differential testing technique that comes with the base fuzzer, \thehuzz{}, to detect vulnerabilities with the RISC-V ISA simulator, \texttt{SPIKE}~\cite{spike} as the reference model.
We evaluate the ability of \toolname{} to detect vulnerabilities by comparing its detection speed with that of \thehuzz{} on seven different vulnerabilities(Table~\ref{tab:b_list_v1}). 
\toolname{} is slower than \thehuzz{} in detecting vulnerability V5 because the vulnerability itself is easy to detect (\thehuzz{} detects it in only 25 tests).
On the other hand, \toolname{} detects all the other vulnerabilities faster than \thehuzz{}, achieving up to $308.89\times$ speedup. 
\toolname{} achieved this highest speedup when detecting vulnerability, V7. 
V7 is a bug in the decode module, which consists of many \texttt{if-else} conditions requiring more exploration than exploitation to detect it. 
Hence, it can be seen that \thehuzz{}, which emphasizes exploitation of the same seeds, takes a long time to detect this bug, whereas \toolname{} detects it up to two orders of magnitude faster.  
Also, the speedup achieved by the MAB algorithms varies with the vulnerabilities, with no single algorithm achieving the highest detection speed for all the vulnerabilities. 
This shows the importance of the compatibility of \toolname{} to support different MAB algorithms, which can be chosen for different use cases.

\subsection{Coverage Analysis}

Fig.~\ref{fig:covPlot} plots the number of branch coverage points achieved by the algorithms used in~\toolname{}
on all benchmarks and compares it with \thehuzz{}. As shown,~\toolname{} outperforms \thehuzz{}.
Furthermore, Fig.~\ref{fig:covRate} shows the coverage speedup and percentage increment in coverage achieved by \toolname{} compared to \thehuzz{} when run for 50,000 tests. 
\toolname{} is at least $2.98\times$ faster than \thehuzz{} on \cva{} and \rc{}. 
This is because \toolname{} minimizes spending time on low-performing seeds by choosing the seeds dynamically based on their performance and resetting them when they stop covering new points. 
This allows \toolname{} to fuzz with more meaningful tests, resulting in at least $0.23\%$ more covered points than \thehuzz{}.  
Moreover, \toolname{}'s ability to explore design space is more evident as the difficulty of covering the points increases. 
For instance, \thehuzz{} achieves the lowest coverage percentage on \cva{}.
On this processor, \toolname{} achieves its highest speedup, $5.38\times$.
On the other hand, \thehuzz{} achieves $>95\%$ coverage on the \boom{} processor, leaving less room for improvement for \toolname{}. 
This results in \toolname{} not achieving higher coverage than \thehuzz{}.

Overall, all the algorithms of \toolname{} detected vulnerabilities faster than \thehuzz{} and achieved more and faster coverage thanks to the efficient balance of exploration and exploitation of the design space by \toolname{}. 
The most recent tool~\cite{chen2023psofuzz} that improves hardware fuzzers using learning models achieved $4.23\times$ and $1.73\times$ speed up in detecting vulnerabilities and achieving coverage, achieving $0.37\%$ more coverage points on average. 
In comparison, \toolname{}: EXP3 is $14.59\times$ and $3.05\times$ faster than \thehuzz{} in detecting vulnerabilities and achieving coverage, achieving $0.68\%$ more coverage points. 
\section{Discussion}\label{sec:discussion}

\noindent\textbf{Other Avenues for MAB Algorithms in Hardware Fuzzers.} In this work, we focused on using MAB algorithms to select seeds to fuzz. Future research can investigate the application of MAB algorithms to other avenues in hardware fuzzers. For instance, most fuzzers choose mutation operators either randomly or following some static probability distribution. This can be improved using MAB algorithms to dynamically explore and exploit the space of mutation operators. MAB algorithms can also be used to decide parameters such as the number of instructions in a test because different numbers of instructions have different impacts on coverage at different times.

\noindent\textbf{Theoretical Analysis.} \toolname{} relies on modification of existing MAB algorithms for hardware fuzzing.
We evaluate its efficacy empirically on a range of processor designs. 
MAB algorithms have rich theoretical analyses, which is absent for our modified algorithms. In the future, we plan to address this by conducting a theoretical analysis of the modified MAB algorithms to understand them thoroughly and possibly devise better MAB algorithms for hardware fuzzing.
\section{Conclusion}\label{sec:conclusion}

Prior works on detecting vulnerabilities in processors using fuzzing have shown reasonable performance, but they rely heavily on static strategies in their algorithms, limiting their efficacy. To address this limitation, we develop a new technique,~\toolname, to make decisions dynamically and balance the exploration of the new test programs and exploitation of the well-performing test programs using MAB algorithms. However, to develop~\toolname, we face challenges related to the compatibility of MAB algorithms and fuzzing. We overcome these challenges by modifying both, the fuzzing flow and MAB algorithms. As a result, the final~\toolname{} formulation is effective, efficient, and agnostic to any hardware fuzzer. Experimental results demonstrate that~\toolname{} achieves up to $308\times$ speedup in detecting vulnerabilities and up to $5\times$ speedup in covering the design space. Although this work focuses on applying MAB algorithms to selecting test programs to fuzz, several other avenues of hardware fuzzers have the potential to improve when optimized with MAB algorithms. These avenues should be explored in the future to improve the verification and security of processors in the face of increasing complexity.

\section{Acknowledgement}
Our research work was partially funded by the US Office of Naval Research (ONR Award \#N00014-18-1-2058), Intel's Scalable Assurance Program, the European Research Council (ERC project HYDRANOS 101055025), Deutsche Forschungsgemeinschaft (DFG) SFB 1119 CROSSING/236615297, and European Union’s Horizon Europe research and innovation program (No. 101070537).
This work does not in any way constitute an Intel endorsement of a product or supplier.
Any opinions, findings, conclusions, or recommendations expressed herein are those of the authors and do not necessarily reflect those of the US Government, Intel, the European Union, or the European Research Council.

\def\bibfont{\footnotesize}
\bibliographystyle{IEEEtran}
\bibliography{main}

\end{document}